\documentclass[prl,aps,twocolumn]{revtex4-1}
\usepackage{bm}
\usepackage{amsmath}

\newcommand{\supersym}{supersymmetry }

\newcommand{\pdel}[2]{\frac{\partial #1}{\partial #2}}

\usepackage[all, warning]{onlyamsmath}
\usepackage{graphicx}
\usepackage{color}

\begin{document}

\title{Sum rule for the partial decay rates of bottom hadrons
based on the dynamical \supersym of the $\bar s$~quark and the $ud$ diquark}%

\author{Taiju Amano$^{1,2}$}%
\author{Daisuke Jido$^{2,1}$}
\email[email: ]{jido@th.phys.titech.ac.jp}
\author{Stefan Leupold$^{3}$}
\affiliation{$^{1}$Department of Physics, Tokyo Metropolitan University, Minami-Osawa, Hachioji, Tokyo, Japan}
\affiliation{$^{2}$Department of Physics, Tokyo Institute of Technology, Ookayama, Meguro-ku, Tokyo,  Japan}
\affiliation{$^{3}$Institutionen f\"or fysik och astronomi, Uppsala universitet, Box 516, S-75120 Uppsala, Sweden}
\date{\today}%

\begin{abstract}
We investigate the weak decays of  $\bar B_{s}^{0}$ and $\Lambda_{b}$  to 
charm hadrons based on the dynamical \supersym between the $\bar s$ quark and the $ud$ diquark. 
We derive a new sum rule relating the decay rates of the processes 
$\bar B_{s}^{0} \to D_{s}^{+} P^{-}$, $\bar B_{s}^{0} \to D_{s}^{*+} P^{-}$ and
$\Lambda_{b} \to \Lambda_{c} P^{-}$, where $P^{-}$ is a negatively charged meson, such as 
$\pi^{-}$ and $K^{-}$. It is found that the observed decay rates satisfy the sum rule very well.
This implies that the \supersym between the $\bar s$ quark and the $ud$ diquark
is also seen in the wavefunctions of the heavy hadrons and suggests that the $ud$ diquark 
can be regarded as a valid effective constituent for heavy hadrons. 
\end{abstract}

\maketitle

Finding fundamental correlations 
is a clue to understand the structure of strongly interacting systems. 
In electron systems the Cooper pair is a key ingredient 
and its condensation leads to superconductivity~\cite{BCS}. 
In nuclear physics,
the nucleon pair correlation is an important object to describe the nuclear structure 
in the interacting boson model~\cite{IBM1,IBM2}. Also the two-neutron correlation can be 
a hint to understand the structure of unstable light nuclei~\cite{dineutron,Migdal73}. 
In hadron physics, the two-quark correlation called diquark has been 
already mentioned in Ref.~\cite{Gell-Mann:1964ewy} when quarks were proposed,
and it can be used as an effective constituent in many-body systems.
The importance of the diquark correlation in hadronic systems was discussed
phenomenologically in~\cite{Anselmino:1992vg,Jaffe:2004ph}.
It is also known that diquark condensation induces color superconductivity 
at high density quark systems~\cite{CSC,Rapp:1997zu}.

The diquark is a colored object that cannot be observed at low energies as an isolated particle 
due to color confinement. Its existence, however, is expected 
as a constituent inside hadrons similar to the constituent quark, 
which is a quasi-particle of the fundamental 
particles and is regarded as an effective building block of hadrons. The role of the diquark in 
the baryon structure has been extensively investigated by diquark pictures,
in which baryons are composed of a diquark and a constituent quark~\cite{Ida66,Lichtenberg67,Goldstein:1979wba,Lichtenberg:1969sxc,Lichtenberg:1982jp,Liu:1983us,Hernandez:2008ej,Lee:2009rt,Jido:2016yuv,Eichmann:2016yit,Kumakawa:2017ffl}.
Light scalar mesons may be described by a configuration of diquark and 
antidiquark~\cite{Jaffe:1976ig,Black:1999yz,Maiani:2004uc,tHooft:2008rus} and their decay properties are 
reproduced reasonably well~\cite{Maiani:2004uc,tHooft:2008rus}.  
Lattice QCD calculations also
have suggested attractive diquark correlations~\cite{Hess98,Alexandrou:2006cq}.

Recently a dynamical \supersym between the $ud$ scalar diquark and the $\bar s$ 
constituent quark has been proposed in Ref.~\cite{Amano:2019jek}.
Both objects have the same color charge~$\bar {\bf 3}$ and same electric charge. 
Phenomenologically they are known to have a similar mass around 500~MeV.
This is a supersymmetry between a boson and a fermion, but not a symmetry 
for fundamental particles, rather a dynamical symmetry 
for quasi particles which are regarded as effective elements of the dynamics like 
the constituent quarks. 
If this \supersym is realized universally in hadronic systems, one may conclude 
the existence of the diquark inside hadrons as seen for the constituent quarks that 
were established from the symmetry arguments of the light hadrons. 
Historically such a dynamical supersymmetry was introduced first by Miyazawa
for mesons and baryons~\cite{Miyazawa} and later applied to the light hadron spectra~\cite{Catto}.
{One can also use holographic QCD to motivate a supersymmetry connecting baryons and mesons \cite{deTeramond:2014asa,Dosch:2015nwa,Nielsen:2018uyn}.}

The \supersym among the scalar $ud$ diquark and the $\bar s$ quark works rather well 
for the hadron spectra~\cite{Amano:2019jek}.  
For instance, combining a bottom quark $b$ with the $ud$ diquark and the $\bar s$ quark,
we have three hadrons $(\bar B_{s}^{0}, \bar B_{s}^{*0}, \Lambda_{b})$, 
which are a spin 0 pseudoscalar meson, a spin 1 vector meson and a spin 1/2 baryon.
The observed masses are found as (5367, 5415, 5620)
in units of MeV, respectively. 
Similarly for the charm quark $c$, we have $(D_{s}^{+}, D_{s}^{*+}, \Lambda_{c})$
and these masses are  (1968, 2112, 2286) in units of MeV.
The symmetry breaking on these hadron masses is about 300 MeV, which is 
as good as the flavor symmetry breaking stemming from the mass difference 
among the light constituent quarks. 

The symmetry among these hadrons may be based on a similar mass
for the $ud$ diquark and the $\bar s$ constituent quark. 
Color electric interactions play the main role for confinement and 
are mainly determined by the masses and color of the interacting particles. 
Because $ud$ diquark and $\bar s$ quark have same color and a similar mass, 
the interactions of the heavy quark with the $ud$ diquark and the $\bar s$ quark must be very similar. 
Possible sources of symmetry breaking are
the mass difference between the $ud$ diquark and the $\bar s$ quark 
and spin dependent forces such as the spin-spin interaction between quarks. 
The former is responsible for the mass difference of the mesons and the baryon, 
while the latter induces the mass difference between pseudoscalar and vector mesons. 

The purpose of this article is to investigate whether this \supersym is realized also
in the wavefunctions in heavy hadrons.  
The symmetry of the wavefunctions can be seen in the decay of the heavy hadrons, 
where the decay rates are expressed by the matrix elements of the
parent and daughter particles with the wavefunctions of the initial and final states. 
For this purpose, we compare the weak decays of $\bar B_{s}^{0}$ into 
$D_{s}^{+}$ and $D_{s}^{*+}$ with those of $\Lambda_{b}$ into $\Lambda_{c}$. 

From now on, 
let us call the $\bar s$ quark and the $ud$ diquark collectively as $\hat \psi$
and consider a spin doublet $\bar s$ and a scalar $ud$ to form a triplet $\hat \psi$ of 
the V(3) \supersym introduced by Miyazawa~\cite{Miyazawa}. We denote 
hadrons composed of the triplet $\hat \psi$ and a heavy quark $h$ collectively 
as $\hat \psi h$. This yields 
$\hat \psi b = (\bar B_{s}^{0}, \bar B_{s}^{*0}, \Lambda_{b})$ 
for the bottom hadrons and $\hat \psi c = (D_{s}^{+}, D_{s}^{*+}, \Lambda_{c})$ 
for the charm hadrons. Hadrons $\hat \psi h$ form a sextet of V(3)$\,\otimes\,$SU(2)
where SU(2) denotes the spin symmetry of the heavy quark.

We consider several weak decay modes in parallel;
pionic decay $\hat \psi b \to \pi^{-} \hat \psi c $,
kaonic decay $\hat \psi b \to K^{-} \hat \psi c $,
$\rho$ mesonic decay $\hat \psi b \to \rho^{-} \hat \psi c $, 
$D$ mesonic decay $\hat \psi b \to D \hat \psi c$,
$D_{s}$ mesonic decay $\hat \psi b \to D_{s} \hat \psi c$,
$D^{*}_{s}$ mesonic decay $\hat \psi b \to D^{*}_{s} \hat \psi c$,
and
semileptonic decay $\hat \psi b \to  \ell^{-} \bar \nu_{\ell}\, \hat \psi c $.
We abbreviate these decay as 
$\hat \psi b \to P \hat \psi c$, where $P$ stands for 
the emitted particles, that is a pion, a kaon, \ldots\
for the mesonic decay
and leptons for the leptonic decay.

{Let us first consider pionic decay $\hat \psi b \to \pi^{-} \hat \psi c $. 
This decay is induced by transition $b \to c W^{-}$
and then either the weak boson $W^{-}$ turns into 
a pion $\pi^{-}$ or $W^{-}$ is absorbed into $\hat \psi$. 
In the former process, the $\bar s $ quark or the $ud$ diquark is a spectator in the weak decay, 
and thus the weak transition of the $b$ quark commonly contributes to the decays of 
$\bar B_{s}^{0}$ and $\Lambda_{b}$ and the wavefunctions of the $\bar s$ quark 
in $\bar B_{s}^{0}$ and of the $ud$ diquark in $\Lambda_{b}$ are responsible 
for the difference of their decay rates.
The latter process involves two particles in the initial state.
Because such a two-body process is known to be strongly suppressed compared 
to one-body processes~\cite{Oset:2016lyh}, we can safely neglect it.
Therefore, the decay process $\hat \psi b \to \pi^{-} \hat \psi c $ is good to investigate 
the supersymmetry in the $\bar s$ and $ud$ wavefunctions. This situation is also true 
for kaonic, $\rho$ mesonic and semileptonic decays.}

\begin{figure}[b]
\vspace{10pt}
\includegraphics[width=0.45\linewidth]{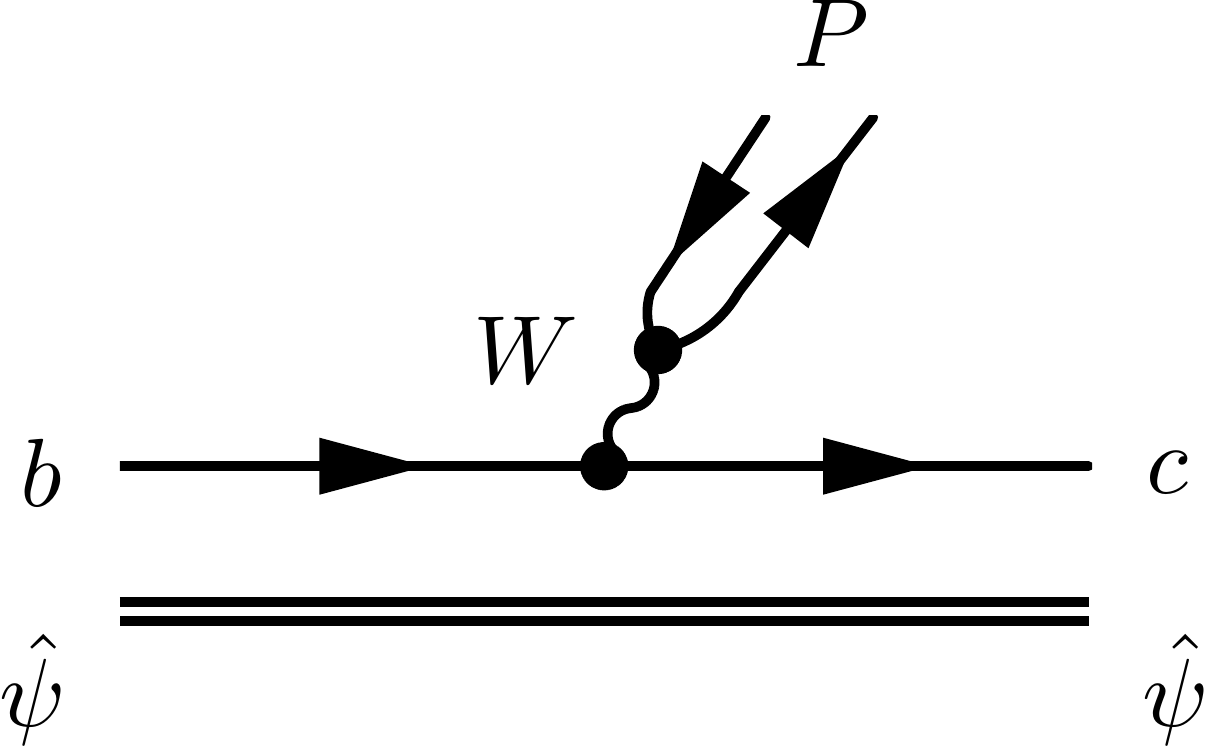}
\quad\quad
\includegraphics[width=0.45\linewidth]{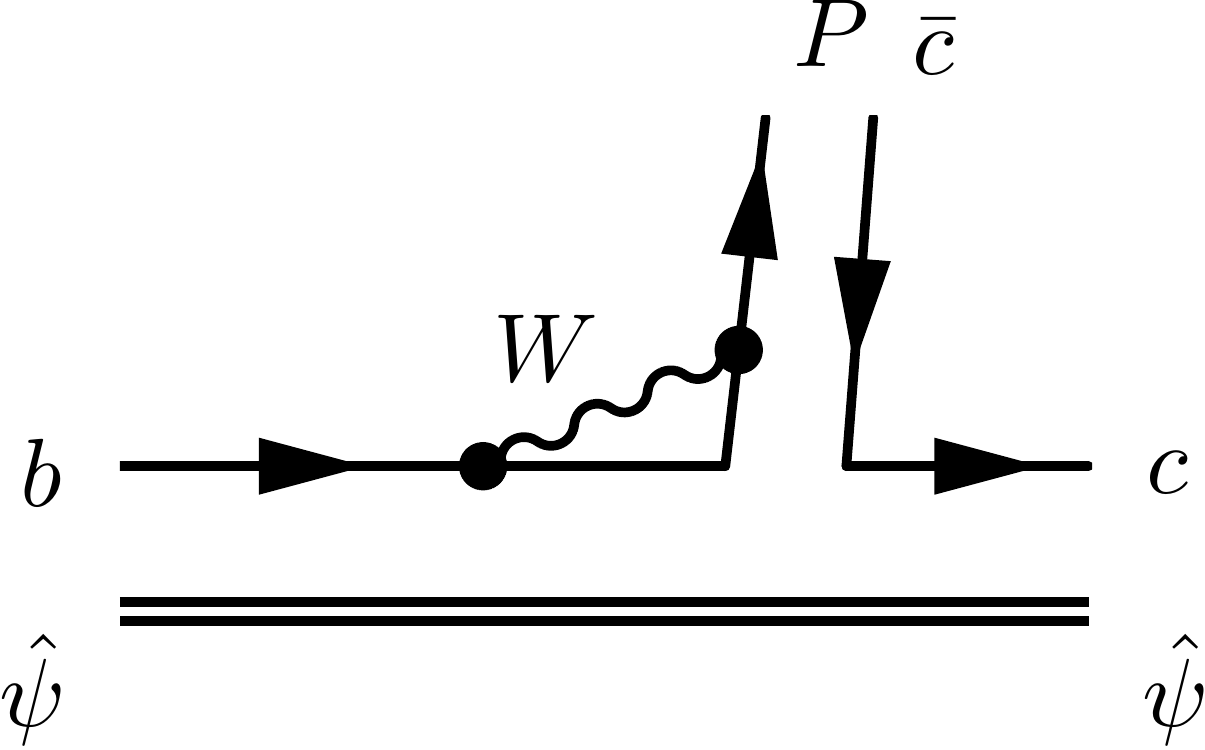}
\\
(a) external $W$-emission
\qquad\quad
(b) horizontal $W$-loop
 \\

\vspace{10pt}
\includegraphics[width=0.45\linewidth]{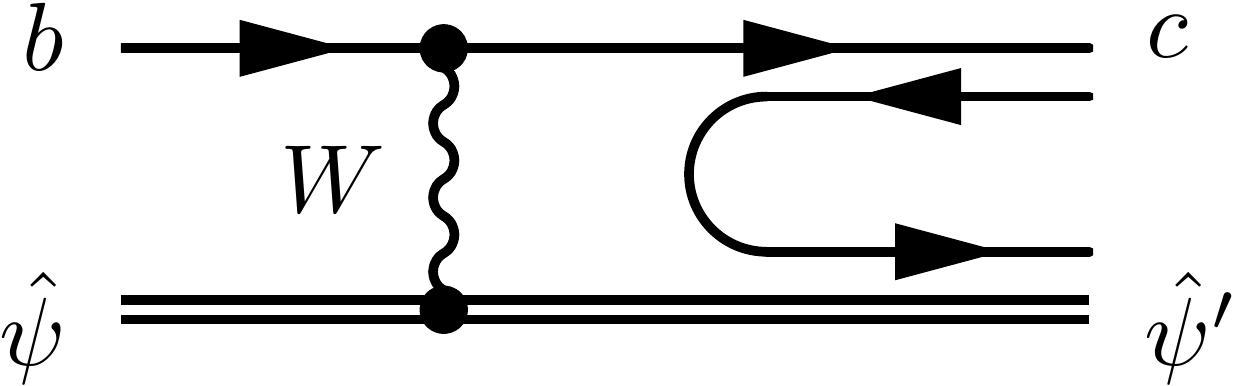}
\quad\quad
\includegraphics[width=0.45\linewidth]{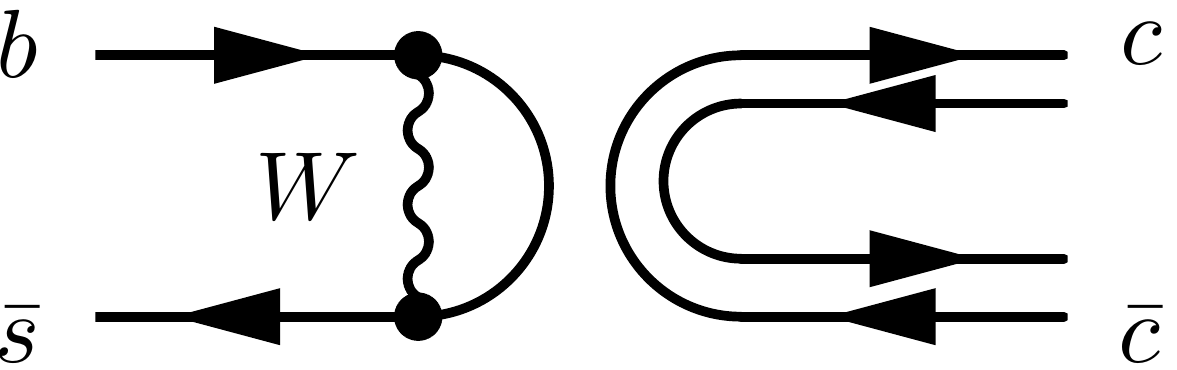}
\\
(c) $W$-exchange
\qquad\qquad\quad
(d) vertical $W$-loop

\caption{Relevant diagrams for the weak decay of $\hat \psi b$ based on the topological 
classification given in Ref.~\cite{WeakPro}. 
}
\label{fig:diagrams}
\end{figure}

{More systematically, we show 
the relevant diagrams of the weak decays $\hat \psi b \to P \hat \psi c$ 
in Fig.~\ref{fig:diagrams} by}
making use of the topological classification of Ref.~\cite{WeakPro}. Based on the supersymmetry we extend
this classification from mesons to baryons and use it for both. 
In the ``external $W$-emission'' diagram (a),
the weak decay is induced by the transition of the $b$ quark to the $c$ quark with 
emitting a meson $P$ directly from the $W$ boson. 
The ``horizontal $W$-loop'' diagram~(b) contains charm quark 
pair creation and contributes to the $D$, $D_{s}$ and $D_{s}^{*}$ mesonic decays.
(The $D$ mesonic decay is doubly Cabibbo suppressed.) 
Also in this diagram, $\hat \psi$ is a spectator.
Diagrams (c) and (d) contribute differently to the $\bar B_{s}^{0}$ and $\Lambda_{b}$.
These two diagrams, however, contain two-body processes.
There are two more diagrams in the classification of Ref.~\cite{WeakPro}:
the internal $W$-emission and the $W$-annihilation diagrams. 
These diagrams are irrelevant for the present calculation, because the former 
diagram does not contain $D_{s}$, $D_{s}^{*}$ nor $\Lambda_{c}$ in the final state
and the latter is only relevant for a charged meson decay. 
In order to explore the diquark ansatz, we did not consider in Fig.~\ref{fig:diagrams} 
the processes in which the $ud$ diquark falls apart during the weak decay. 

The decays of $\hat \psi b$ can be calculated from diagrams (a) and (b), 
in which $\hat \psi$ can be regarded as a spectator of the decay process. 
The effective Hamiltonian that we consider here for the transition $b$ to $c$ reads
\begin{equation}
   {\cal H}_{W}^{L} 
   = \bar c \gamma^{\mu} (A + B \gamma_{5}) b P_{\mu}  \equiv J_{h}^{\mu} P_{\mu},
   \label{eq:effH}
\end{equation}
where $P_{\mu}$ is the weak current for each weak process, such as
$P^{\mu} = \partial^{\mu} \pi^{\dagger}$ for the pionic decay, 
$P^{\mu} = \rho^{\mu \dagger}$ for the $\rho$ mesonic decay, and
$P^{\mu} =  \bar \ell  \gamma^{\mu} ( 1- \gamma_{5}) \nu_{\ell}$ for the leptonic decay.
The effective coupling strengths $A$ and $B$ in the current $J_{h}^{\mu}$
depend on the weak process specified by $P^{\mu}$, but do {\it not} depend on whether the 
spectator is a $ud$ diquark or an $\bar s$ quark. Here the \supersym 
enters. 

The decay rate of a bottom hadron $\hat \psi b$ (with mass $M_{\hat\psi b}$) to a charm hadron $\hat \psi c$ 
(mass $M_{\hat\psi c}$) by emitting particle(s) $P$ is calculated as 
\begin{equation}
   \Gamma = \frac{1}{2M_{\hat\psi b}} 
   \int \sum_{\textrm{spin}} |{\cal M}_{h}^{\mu} {\cal M}_{P\mu}|^{2} d\Phi_{f} \label{eq:Gamma}
\end{equation}
with the phase space element of the final states $d\Phi_{f} \equiv (2\pi)^{4} \delta^{4}(q - \sum_{i} p_{i}) \prod_{i} 
\frac{d^{3}p_{i}}{2E_{i}(2\pi)^{3}}$.  
Spin average of the 
initial state and spin summation of the final states are taken. 
The matrix elements ${\cal M}_{h}^{\mu}$ and ${\cal M}_{P}^{\mu}$ are 
defined by
\begin{equation}
{\cal M}_{h}^{\mu} = \langle \hat\psi c | J_{h}^{\mu} | \hat \psi b \rangle, \qquad
{\cal M}_{P}^{\mu} = \langle P | P^{\mu} | 0\rangle. \label{eq:matele}
\end{equation}
The latter matrix element ${\cal M}_{P}^{\mu}$ describes the 
particle emission during the transition and is common for the process 
$\hat \psi b \to P \hat \psi c$, irrespective of the choice of the triplet member from $\hat \psi$ and irrespective of 
the spin orientations of the heavy quarks. 
For two-body decays in the rest frame, 
the decay rate~\eqref{eq:Gamma} is written as
\begin{equation}
   \Gamma = \underbrace{\frac{p_\textrm{c.m.}}{32\pi^{2}M_{\hat \psi b}^{2}}}_{\equiv K} 
   \int \sum_{\textrm{spin}} |{\cal M}_{h}^{\mu} {\cal M}_{P\mu}|^{2} d\Omega, \label{eq:GammaTwo}
\end{equation}
with the center of mass momentum of the final states 
$p_\textrm{c.m.}= \{(M_{\hat \psi b}^{2}-(M_{\hat \psi c}+m)^{2})(M_{\hat \psi b}^{2}-(M_{\hat \psi c}-m)^{2})\}^{1/2}/(2M_{\hat \psi b})$ 
where $m$ denotes the mass of particle $P$. 
Thanks to the symmetry of the masses in the same multiplet,
the mass of the decaying hadron and the phase space of the final states are also 
the same in each decay mode specified by $P$.

Because the $\bar s$ quark and the $ud$ diquark can be regarded as spectators
in the transition, the hadronic matrix element ${\cal M}_{h}^{\mu}$ can be evaluated 
in terms of the heavy quark states. Specifying the quark spins, we write
the matrix element of the hadronic current for 
the bottom and charm quarks with spin $\alpha$ and $\beta$ as
\begin{equation}
{\cal M}_{\alpha\beta}^{\mu}
   = \langle c^{(\beta)}  |  J_{h}^{\mu} | b^{(\alpha)} \rangle 
= \bar u_{c}^{(\beta)} \gamma^{\mu} (A + B \gamma_{5}) u_{b}^{(\alpha)}. \label{eq:matrix}
\end{equation}
Under the assumption that the wavefunctions are the same due to the supersymmetry, 
the matrix elements ${\cal M}_{\alpha\beta}^{\mu}$ appear commonly in the 
calculations of each decay mode.

The spin of the heavy baryon $\Lambda_{h}$ and the heavy quark coincide 
thanks to the spinless diquark. 
Thus the spin wavefunctions of the heavy baryon spin doublet 
$\Lambda_{h}^{(1)}$ and $\Lambda_{h}^{(2)}$ are given by 
$(ud) h^{(1)}$ and $(ud) h^{(2)}$,
respectively. 
For the decay rate of an unpolarized $\Lambda_{b}$ to $\Lambda_{c}$,
we take a spin average of the initial $\Lambda_{b}$ and sum up all of the spin states 
of the final $\Lambda_{c}$:
\begin{align}
  \lefteqn{ \sum_{\textrm{spin}} ({\cal M}_{\Lambda_{c}}^{\mu})^{*} 
  {\cal M}_{\Lambda_{c}}^{\nu}=}& \nonumber \\
   & \tfrac12 ({\cal M}_{11}^{\mu*} {\cal M}_{11}^{\nu}+{\cal M}_{22}^{\mu*} {\cal M}_{22}^{\nu}
   + {\cal M}_{12}^{\mu*} {\cal M}_{12}^{\nu} + {\cal M}_{21}^{\mu*} {\cal M}_{21}^{\nu}).
   \label{eq:bar}
\end{align}

The spin configuration of a pseudoscalar meson composed of 
a heavy quark $h$ and an $\bar s$ quark
is given by 
$\frac{1}{\sqrt 2} (\bar s^{(1)} h^{(1)} + \bar s^{(2)} h^{(2)})$.
For the weak decay of the pseudoscalar $\bar B_{s}^{0}$, 
the spin of the $\bar s$ quark does not change in the decay 
as it is a spectator. 
The hadronic part of the decay amplitude of $\bar B_{s}^{0}$ to the pseudoscalar $D_{s}^{+}$
is calculated as
\begin{align}
  \lefteqn{{\cal M}_{D_{s}^{+}}^{\mu} 
  = \langle D_{s}^{+}| J_{h}^{\mu}| B_{s}^{0}\rangle } \nonumber \\
  &=\langle \tfrac{1}{\sqrt 2} (\bar s^{(1)} c^{(1)} + \bar s^{(2)} c^{(2)})
  |J_{h}^{\mu}| \tfrac{1}{\sqrt 2} (\bar s^{(1)} b^{(1)} + \bar s^{(2)} b^{(2)})\rangle \nonumber \\
  &= \tfrac12 ({\cal M}^{\mu}_{11} + {\cal M}^{\mu}_{22}),
\label{eq:ps-me}
\end{align}
where we have used the orthogonality of the states having different spin for the $\bar s$ quark.
This implies that the weak decay of $\bar B_{s}^{0}$ to $D_{s}^{+}$ has only
the spin non-flip amplitude~$A$. 
The square of the amplitude is given by
\begin{align}
   \lefteqn{  ({\cal M}_{D_{s}^{+}}^{\mu})^{*} {\cal M}_{D_{s}^{+}}^{\nu}=} &
   \nonumber \\ 
    & \tfrac14({\cal M}_{11}^{\mu*} {\cal M}_{11}^{\nu}+{\cal M}_{22}^{\mu*} {\cal M}_{22}^{\nu}
   + {\cal M}_{11}^{\mu*} {\cal M}_{22}^{\nu} + {\cal M}_{22}^{\mu*} {\cal M}_{11}^{\nu}).
   \label{eq:ps}
\end{align}

The spin configurations for the vector mesons $D_{s}^{*+}$ with $s_{z} = +1, 0, -1$ are 
given by $\bar s^{(2)} h^{(1)}$, $\tfrac{1}{\sqrt 2} (\bar s^{(1)} h^{(1)} - \bar s^{(2)} h^{(2)})$
and $\bar s^{(1)} h^{(2)}$,
respectively.
In analogy to Eq.~\eqref{eq:ps-me}, 
we calculate
the decay amplitudes of the pseudoscalar $\bar B_{s}^{0}$ to the vector $D_{s}^{*+}$. 
Summing up the spin of $D_{s}^{*+}$
in the final state, we obtain 
\begin{align}
  \lefteqn{ \sum_{\textrm{spin}} ({\cal M}_{D_{s}^{*+}}^{\mu})^{*} 
  {\cal M}_{D_{s}^{*+}}^{\nu}= 
  \tfrac12 {\cal M}_{21}^{\mu*} {\cal M}_{21}^{\nu}
  +  \tfrac12 {\cal M}_{12}^{\mu*} {\cal M}_{12}^{\nu}
  }& \nonumber \\
   & 
  + \tfrac14 ( {\cal M}_{11}^{\mu*} {\cal M}_{11}^{\nu}
  + {\cal M}_{22}^{\mu*} {\cal M}_{22}^{\nu}
  - {\cal M}_{11}^{\mu*} {\cal M}_{22}^{\nu} - {\cal M}_{22}^{\mu*} {\cal M}_{11}^{\nu}).
  \label{eq:vec}
\end{align}

The heavy hadrons
 $(\bar B_{s}^{0},\Lambda_{b})$ and $(D_{s}^{+}, D_{s}^{*+}, \Lambda_{c})$ are
in the same multiplets, respectively, and the V(3) \supersym demands the kinematical factors
of these decays to be the same. In addition, if the wavefunctions of the heavy hadrons are the same 
in each multiplet, the hadronic matrix elements can be calculated commonly using the 
amplitude~\eqref{eq:matrix}. Under these conditions, we find that
the sum of Eqs.~\eqref{eq:ps} and \eqref{eq:vec} coincides with Eq.~\eqref{eq:bar}.
This implies that we have a sum rule for the decay probabilities 
of $\bar B_{s}^{0}$ and  $\Lambda_{b}$ as
\begin{equation}
  \Gamma_{\bar B_{s}^{0} \to D_{s}^{+}} + \Gamma_{\bar B_{s}^{0}\to D_{s}^{*+}}
  =   \Gamma_{\Lambda_{b}\to \Lambda_{c}}. \label{eq:SR}
\end{equation}
With this sum rule, we can check the symmetry of the wavefunctions for the heavy 
hadrons $\hat \psi h$. 


We will examine whether the sum rule~\eqref{eq:SR} agrees with experimental observations and we will derive predictions 
for partial decay rates that have not been measured yet. 
The experimental data collected by the Particle Data Group (PDG)~\cite{PDG} are summarized in Table~\ref{tab:data}, where
the partial decay rates are evaluated in units of $10^{9}/\textrm{s}$ using the central values 
of the mean life of the decaying particle and the branching fraction of the corresponding 
decay mode. For $\bar B_{s}^{0}$, we use the average of the mean lives of the heavy and
light $CP$ eigenstates~\cite{HFLAV:2019otj}. 
Although the branching fractions for $B_{s}^{0}$ are provided by the PDG, we use them for $\bar B_{s}^{0}$ 
since $CP$ violation is very small. 

First of all, it is very interesting to note that for each decay mode
the partial decay rates of the $\bar B_{s}^{0}$ meson and the $\Lambda_{b}$ baryon 
have the same order of magnitude. This can be interpreted already as a consequence of the \supersym 
between the $\bar s$ quark and the $ud$ diquark.

For the decays $\hat\psi b \to \pi^{-} \hat \psi c$, 
the sum of decay rates of $\bar B_{s}^{0} \to \pi^{-} D_{s}^{+}$ 
and $\to \pi^{-} D_{s}^{*+}$ yields $(3.3 \pm 0.4) \times 10^{9} / \textrm{s}$,
while the decay rate $\Lambda_{b} \to \pi^{-} \Lambda_{c}$
is $(3.3 \pm 0.3) \times 10^{9} / \textrm{s}$. 
Thus, the sum rule~\eqref{eq:SR} is satisfied extremely well.  

Next, we discuss the sum rule~\eqref{eq:SR} for $\hat\psi b \to K^{-} \hat \psi c$. 
Unfortunately, present experiments provide only the branching fraction of 
the $B_{s}^{0} \to K^{\pm} D_{s}^{(*)\mp}$ decay, i.e.\ one cannot discriminate between
$B_{s}^{0} \to K^{+} D_{s}^{(*)-}$ and $B_{s}^{0} \to K^{-} D_{s}^{(*)+}$. 
Therefore we consider the kaonic decay fractions for $\bar B_{s}^{0}$
as upper limits. 
The sum of the decay rates of $\bar B_{s}^{0}$ to $K^{\mp} D_{s}^{\pm}$ 
and $K^{\mp} D_{s}^{*\pm}$ is found to be $(2.37 \pm 0.26) \times 10^{8} / \textrm{s}$,
while the decay rate of $\Lambda_{b}$ to $K^{-} \Lambda_{c}$
is observed as $(2.44 \pm 0.020) \times 10^{8} / \textrm{s}$. 
The sum rule may work well. 

For the $\rho$, $D^{-}$ and $D_{s}$ mesonic decays, 
one of the branching fractions has not been measured yet. 
Assuming the sum rule \eqref{eq:SR},
we can predict the partial decay rates of these missing decay modes.
The predicted values are shown as values in square brackets in Table~\ref{tab:data}. 
It will be very interesting to see if future measurements of the branching rates of 
presently missing decays will confirm the validity of our sum rule \eqref{eq:SR}. 
The decay branching fraction of 
$\bar B_{s}^{0} \to D_{s}^{-} D_{s}^{*+} + D_{s}^{*-} D_{s}^{+} $ 
has been observed as $(1.39\pm{0.17})\times 10^{-2}$,
which corresponds to $(9.17\pm1.12) \times 10^{9}/\textrm{s}$ for the partial decay rate. 
Using the partial decay rate of $\bar B_{s}^{0} \to D_{s}^{-} D_{s}^{*+}  $ 
obtained from the sum rule, we estimate the partial decay rate of
$\bar B_{s}^{0} \to D_{s}^{-} D_{s}^{*+}$ as $(4.6 \pm 0.8) \times 10^{9}/\textrm{s}$.
Using the sum rule again, we can predict the partial decay rate
of $\Lambda_{b} \to D_{s}^{-} \Lambda_{c}$ as $(14.1\pm 1.9) \times 10^{9}/\textrm{s}$.

For the semileptonic decays, 
exclusive measurements exist only for the baryon case. For the $\bar B_{s}^{0}$ decays they have not been 
performed yet. But the three inclusive decay modes collected in Table~\ref{tab:data} have a similar magnitude to the 
baryon decay rate.
This may be a consequence of the \supersym between $\bar s $ and $ud$. 
In order to confirm the sum rule for the semileptonic decays, exclusive observations 
are strongly desired.

\begin{table}
\caption{Weak decay modes of $\bar B_{s}^{0}$ and $\Lambda_{b}$ and 
the corresponding branching fractions and rates.
The partial decay rates $\Gamma_{i}$ are shown in units of
$10^{9}/\textrm{s}$ and are evaluated using the central values of 
the observed mean life and branching fraction. 
The values of the partial decay rates in square brackets are predictions based on 
the sum rule \eqref{eq:SR}.
The value of the observed mean life of $\bar B_{s}^{0}$ is 
$(1.515 \pm 0.004) \times 10^{-12}$s, which is the average mean life 
of the heavy and light $CP$ eigenstates, 
and that of $\Lambda_{b}$ is
$(1.471 \pm 0.009) \times 10^{-12}$s. 
The charge of the kaonic decay of $\bar B_{s}^{0}$ cannot be 
discriminated due to the $\bar B_{s}^{0}$-$B_{s}^{0}$ mixing.
The data are taken from~\cite{PDG}.
The original experiments are found in Refs.~\cite{LHCb:2012wdi,Belle:2008ezn,CDF:2006hob,Belle:2010ldr,LHCb:2015jtt,LHCb:2014scu,LHCb:2013sad,Belle:2012tsw,CDF:2012xmd,LHCb:2016ggs,Belle:2015ftp,ALEPH:1995npv,DELPHI:1992iji,OPAL:1992zsd,LHCb:2014ofc,LHCb:2014yin,CDF:2006vnk,ALEPH:1997ake,DELPHI:1995dtx}.}
\label{tab:data}
\begin{tabular}{l rr}
\hline\hline
  & \multicolumn{1}{c}{branching fraction} & $\Gamma_{i}$ $[10^{9}/\textrm{s}]$ \\
\hline
pionic decay &&\\
$\bar B_{s}^{0} \to \pi^{-} D_{s}^{+}$ & $(3.00\pm{0.23})\times 10^{-3}$ & $1.98 \pm 0.15$\\
$\bar B_{s}^{0} \to \pi^{-} D_{s}^{*+}$ & $(2.0\pm{0.5})\times 10^{-3}$ &$1.3 \pm 0.3$\\
 sum &   & $[3.3 \pm 0.4]$ \\
$\Lambda_{b} \to \pi^{-} \Lambda_{c}$ &$(4.9\pm{0.4})\times 10^{-3}$&$3.3 \pm 0.3$\\
\hline
kaonic decay \\  
$\bar B_{s}^{0} \to K^{-} D_{s}^{+}$ & $< (2.27\pm{0.19})\times 10^{-4}$ & $< (0.150 \pm 0.013)$\\
$\bar B_{s}^{0} \to K^{-} D_{s}^{*+}$ & $< (1.33\pm{0.35})\times 10^{-4}$ &$< (0.088 \pm 0.023)$\\
sum & & $[< (0.237 \pm 0.026)]$ \\
$\Lambda_{b} \to K^{-} \Lambda_{c}$ &$(3.59\pm{0.30})\times 10^{-4}$&$0.244 \pm 0.020 $\\
\hline
$\rho$ mesonic decay \\  
$\bar B_{s}^{0} \to \rho^{-} D_{s}^{+}$ & $(6.9\pm{1.4})\times 10^{-3}$ & $4.6 \pm 0.9$ \\
$\bar B_{s}^{0} \to \rho^{-} D_{s}^{*+}$ & $(9.6\pm{2.1})\times 10^{-3}$ & $6.3 \pm 1.4$\\
$\Lambda_{b} \to \rho^{-} \Lambda_{c}$ & $[(16.0 \pm 2.4)\times 10^{-3}]$ & $[10.9 \pm 1.7]$ \\
\hline
$D$ mesonic decay \\  
$\bar B_{s}^{0} \to D^{-} D_{s}^{+}$ & $(2.8\pm{0.5})\times 10^{-4}$ & $0.18 \pm 0.03$\\
$\bar B_{s}^{0} \to D^{-} D_{s}^{*+}$ & $[(1.9 \pm 0.8)\times 10^{-4}]$ &$[0.13 \pm 0.05]$ \\
$\Lambda_{b} \to D^{-} \Lambda_{c}$ &$(4.6\pm{0.6})\times 10^{-4}$& $0.31 \pm 0.04 $\\
\hline
$D_{s}$ mesonic decay \\  
$\bar B_{s}^{0} \to D_{s}^{-} D_{s}^{+}$ & $(4.4\pm{0.5})\times 10^{-3}$ & $2.9 \pm 0.3$\\
$\bar B_{s}^{0} \to D_{s}^{-} D_{s}^{*+} $ & $[(6.9 \pm 1.1)\times 10^{-3}]$  & $[4.6 \pm 0.8]$\\
$\Lambda_{b} \to D_{s}^{-} \Lambda_{c}$ &$(1.10\pm{0.10})\times 10^{-2}$& $7.5 \pm 0.7$\\
\hline
$D^{*}_{s}$ mesonic decay \\  
$\bar B_{s}^{0} \to  D_{s}^{*-} D_{s}^{+}  $ & $[(0.70 \pm 0.20)\times 10^{-2}]$ & $[4.6 \pm 1.4]$\\
$\bar B_{s}^{0} \to D_{s}^{*-} D_{s}^{*+}$ & $(1.44\pm{0.21})\times 10^{-2}$ & $9.5 \pm 1.4$\\
$\Lambda_{b} \to D_{s}^{*-} \Lambda_{c}$ & $[(2.07 \pm 0.28)\times 10^{-2}]$  &$[14.1 \pm 1.9]$ \\
\hline
semileptonic decay \\  
$\bar B_{s}^{0} \to \ell^{-} \bar\nu_{\ell} D_{s}^{+}+X$ & $(8.1\pm{1.3})\times 10^{-2}$ & $53\pm 9$\\
$\bar B_{s}^{0} \to \ell^{-} \bar\nu_{\ell} D_{s}^{*+} + X$ & $(5.4\pm{1.1})\times 10^{-2}$& $36\pm7$\\
$\Lambda_{b} \to \ell^{-} \bar\nu_{\ell} \Lambda_{c}+X$ &$(10.9\pm{2.2})\times 10^{-2}$& $74\pm15$\\
$\Lambda_{b} \to \ell^{-} \bar\nu_{\ell} \Lambda_{c}$ &$(6.2^{+1.4}_{-1.3})\times 10^{-2}$& $42^{+10}_{-9}$\\
\hline\hline
\end{tabular}
\end{table}

It is interesting to estimate the magnitude of symmetry breaking 
of the sum rule~\eqref{eq:SR} coming from the kinematical factor $K$ of Eq.~\eqref{eq:GammaTwo}.
This factor is a function of $M_{\hat \psi b}$, $M_{\hat \psi c}$ and $m$. 
The observed heavy hadron masses deviate from the symmetric mass~$M_{h}$. 
The latter is given by a spin average
$M_{b} = (M_{\bar B_{s}^{0}} + 3 M_{B_{s}^{*0}} + 2 M_{\Lambda_{b}})/6$ and similar for the charm sector.
Numerically one obtains $M_{b}=5475$ MeV and $M_{c}=2146$ MeV. 
The deviation of the kinematical factor $K$ from the symmetry limit can be evaluated as
\begin{equation}
  \frac{K(M_{\hat \psi b}, M_{\hat \psi c}, m)}{K(M_{b}, M_{c}, m)}  
  \approx 1 + \pdel{\log K}{M_{\hat \psi b}} \delta m_{b}
  + \pdel{\log K}{M_{\hat \psi c}} \delta m_{c} , \label{eq:dev}
\end{equation}
where $\delta m_{b}$ and $\delta m_{c}$ are the deviations of the bottom and charm
hadron masses from their symmetric mass, respectively. Evaluating Eq.~\eqref{eq:dev}
using the observed masses, we find that the deviation of the kinematical factor 
from the symmetric limit is 5 \% at most for these decay modes. 
Therefore, the fact that the sum rule works very well for the observed weak decay 
processes implies that the wavefunctions for the heavy hadrons $\hat \psi h$
have also good symmetry stemming from the \supersym between the $\bar s$ constituent quark 
and the $ud$ diquark.

In conclusion, based on the \supersym between the $\bar s$ quark 
and the $ud$ scalar diquark, we have derived a sum rule for the weak
transition rates of the bottom $\bar B_{s}^{0}$ meson and $\Lambda_{b}$ baryon
to charm hadrons. The sum rule is well satisfied by the observed weak decays 
for pionic and kaonic decay modes.
This implies that the $ud$ scalar diquark behaves as a quasi-particles inside
of the $\Lambda_{b}$ and  $\Lambda_{c}$ baryons like the $\bar s$ quark
in heavy mesons and can be a clue for the nature of the diquark. We have also 
predicted from the sum rule several weak decay rates of $\bar B_{s}^{0}$ and $\Lambda_{b}$
that have not been observed yet. If these missing decay modes
are observed in future experiments, they can give us further support for the importance 
of the diquark correlation.

\begin{acknowledgments}
This work was partially supported by the Grant-in-Aid for Scientific Research (Grant Numbers JP17K05449 and JP21K03530) from the Japan Society for the Promotion of Science and by the Swedish Research Council (Vetenskapsr\aa det) (grant number 2019-04303).
\end{acknowledgments}

\end{document}